\documentclass[a4paper,11pt]{article}
\pdfoutput=1
\usepackage{jcappub}

\usepackage{array}
\usepackage{mathtools}

\usepackage{etoolbox}

\title{Dark matter from primordial quantum information}
 
\author[a]{C\'esar G\'omez,} 
\emailAdd{cesar.gomez@uam.es}

\author[b,c]{Raul Jimenez} 
\emailAdd{raul.jimenez@icc.ub.edu}

\affiliation{Instituto de F\'{i}sica Te\'orica UAM-CSIC, Universidad Aut\'onoma de Madrid, Cantoblanco, 28049 Madrid, Spain.}
\affiliation[b]{ICC, University of Barcelona, Marti i Franques 1, 08028 Barcelona, Spain.}
\affiliation[c]{ICREA, Pg. Lluis Companys 23, Barcelona, E-08010, Spain.}

\date{\today}

\abstract{
We suggest a general relation between the position of the cosmic microwave background temperature power spectrum peaks and the inflationary slow roll parameter $\epsilon$. This relation is based on interpreting the variable setting the position of the peaks as the quantum distance between the end of inflation and recombination. This distance is determined by the primordial cosmological Fisher information introduced in \cite{GomezJimenez}.  The observational constraints set by cosmic microwave background temperature data lead to a very stringent prediction for the value of the tensor-to-scalar ratio: $r=0.01 \pm 0.002$. Future polarization data of the cosmic microwave background  should be able to measure this signal and corroborate or discard our model.
}

\begin{document} 
\maketitle

\section{Introduction}

In a recent paper \cite{GomezJimenez} we have put forward a new approach to describe early Universe cosmology, the inflationary period, based on information theory. In a nutshell, the key idea lies in representing the power spectrum of the primordial fluctuations of the curvature in terms of a cosmological quantum Fisher information function. 

The logic underlying this approach is easy to understand. Primordial quantum fluctuations are generated by time fluctuations representing the local time delays at the end of inflation. Therefore, the corresponding power spectrum \cite{Mukhanov, Hawking} can be simply interpreted as the variance of the quantum time estimator, something that we can define using the corresponding quantum Fisher information \footnote{For some basic material underlying this point of view see \cite{Gomez} and references therein.}.  

In General Relativity, any physical clock should be associated with a dynamical field. This can be thought as a natural consequence of general covariance. In the case of inflationary Cosmology, the inflaton plays the natural role of a classical clock. Time is provided by the rolling of the scalar field (or the dominant field if it is multi-field inflation) from its initial state until the end of inflation. Quantum fluctuations in a de Sitter background i.e. in a primordial expanding Universe, lead to the well known features of the power spectrum of curvature fluctuations that provide the seeds for structure formation. The basic point of our approach lies in using as basic quantum field to define the quantum clock, the standard self-adjoint quantum estimator operator of time. Using quantum estimation theory, we infer the quantum uncertainty of this quantum estimator i.e. the power spectrum. In order to do this, we need to define a cosmological  quantum Fisher function encoding the information about the primordial expanding de Sitter Universe.

To define this cosmological Fisher information we use two basic ingredients. On one side a quasi de Sitter model of the inflationary period from which we extract the relative entanglement entropy that roughly measures the quantum distance between different moments of the evolution of the Universe during inflation. Once we have defined a relative entanglement entropy as associated with the inflationary period, we define the quantum Fisher information as the second derivative of the relative entanglement entropy, using as a reference density matrix a thermal one defined using the Gibbons-Hawking temperature \cite{Gibbons}. The so defined quantum Fisher function depends on typical slow roll parameters. In particular, the basic Fisher function associated with the physical cosmological time $t$ is given by $F = 2b \epsilon^2$.  

In order to fix the value of $b$ we need to identify the observed power spectrum for curvature fluctuations 
\begin{equation}
\Delta \sim \frac{H^2}{8 \pi^2\epsilon}
\end{equation} 
in natural Planck units,
with $\frac{\dot \phi^2}{M_P^2 F}$. This correspondence is based on identifying the power spectrum for the time quantum estimator operator $\hat \delta(t)$ as $\frac{1}{F}$ and to define the 
curvature fluctuation operator as $\dot \phi \frac{\hat \delta(t)}{M_P}$ for $\phi$ the inflaton field ( see \cite{sky} for details ). Note that in this construction we are defining the power spectrum for the quantum estimator $\hat \delta(t)$ as $\frac{1}{F(t)}$ and not as  the statistical variance of time at the Cramer-Rao bound given by ${\rm Var}(t) = \frac{1}{N F(t)}$ for $N$ the number of measurements used to estimate time. In this sense we put in correspondence the observed power spectrum at horizon exit with the power spectrum of the quantum estimator operator and not with the statistical variance.

Irrespectively of the intrinsic value of this approach, it is natural to try to go a bit further and to see if this information picture of inflation can sheds some light on the way the primordial fluctuations are encoded in the cosmic micorwave background (CMB) power spectrum. Obviously, this is a much more difficult issue since the way primordial fluctuations are imprinted in the the CMB spectrum involves a rather complicated dynamics that depends on the peculiar distribution of the energy density after reheating. 

There are two main experimental results that point at inflation as the correct picture for the early universe: the existence of super-horizon fluctuations~\cite{verde} and the red tilt in the power spectrum of initial fluctuations~\cite{wmap,Planck18}. In order to reproduce the amplitudes of the peaks of the CMB power spectrum, it is necessary to postulate dark matter as a component of the energy budget of the Universe. 

Furthermore,  the observed spectrum of acoustic peaks in the CMB power spectrum, provides additional support for inflation. Crucial to the existence of these peaks is the so called phase coherence argument \cite{Albrecht,Dodelson}. In brief,  what this argument implies is that all the Fourier modes, with the same amplitude, generated during inflation, re-enter the horizon with the same phase i.e., simultaneously. This means that only cosine modes enter into the constructive interference creating the acoustic modes. The physics underlying this phenomena i.e., how phase decoherence disappears before modes reenter the horizon, is a very important smoking gun of the inflationary picture in the regime of multipoles where the acoustic peaks appear. Before recombination, the origin of these peaks depends on the competition between gravitational attraction and radiation pressure. To have these effects under control it is important to postulate a Frieman-Roberstson-Walker metric  in addition to the hydrodynamics describing the baryon photon fluid \footnote{Note that potential effects of phase decoherence can appear at very high multipoles.} 

\section{Results}

Using the standard conformal time, the location of the peaks is determined by the argument of the cosine Fourier modes. The relevant parameter~\cite{cmbslow} is
\begin{equation}\label{one}
\rho = \frac{1}{\eta_0} \int_0^{\eta_{rec}} c_s(\eta) d\eta
\end{equation}
for $c_s$ the speed of sound and $\eta_{rec}$ the recombination time. The location of the temperature CMB peaks can be expressed in terms of $\rho$ as
\begin{equation}\label{peaks}
l_n = \pi \rho^{-1} (n - \frac{1}{8}),
\end{equation} 
which is eq.~(101) in Ref.~\cite{cmbslow}. This equation has corrections of $\sim 10\%$ depending of whether the peaks are odd or even, but this level of accuracy will not be relevant to our argument.

 Let us rewrite (\ref{one}) as
\begin{equation}
\rho = \frac{1}{\eta_{rec}} \int_0^{\eta_{rec}} \tilde c_s(\eta) d\eta
\end{equation}
with 
\begin{equation}
\tilde c_s \equiv \frac{c_s \eta_{rec}}{\eta_0}.
\end{equation}
This simple rewriting is introduced in order to think of the parameter $\rho$ as a time average. For reasons that will become clear below let us denote this time average $<\Delta(E)>$ with 
\begin{equation}
\Delta(E) = \frac{c_s \eta_{rec}}{\eta_0}.
\end{equation} 

With this simple rewriting the reader can immediately guess what we are after.  Thinking in simple quantum mechanical terms what we are trying to do is to identify
\begin{equation}
\int_0^{\eta_{rec}} \tilde c_s(\eta) d\eta
\end{equation}
with the length in quantum Hilbert space of a path of states that go from some initial condition at $\eta_0$ into the state at recombination time $\eta_{rec}$. The induced Fubini-Studi metric defining this length is again the quantum Fisher function that roughly defines the variance of the energy of the initial state \cite{Aharonov, paris}. 

Of course, this metric depends on what is the Hamiltonian inducing the evolution. This Hamiltonian contains all the information about the very complicated dynamics between reentering the horizon and the recombination time. We do not have these data under control but we can consider as a first approximation a very crude assumption, namely that the evolution is determined by a unique hermitian Hamiltonian. Taking this assumption, we can ask ourselves for the most natural guess for the quantum Fisher function defining $\rho$ as a length in Hilbert space i.e. 
\begin{equation}
\rho \sim \frac{1}{\eta_{rec}} \int_{\eta_0}^{\eta_{rec}} \sqrt{F} d \eta.
\end{equation}
Recall that this length simply informs us about the quantum distinguishability between the quantum state at the end of the inflationary period and the state at recombination time. The distinguishability of these states should be totally determined by what happens during the time of recombination since the state is actually not changing during the long period between the exit of the horizon and reentering. 

Now we shall make our basic assumption. Namely, we will take as the natural, model independent estimate for the quantum Fisher function, the one we have associated {\bf with time} with the inflationary period itself $F= 16\pi^2\epsilon^2$; the numerical factor $b=8\pi^2$ is set by imposing numerical agreement with the standard power spectrum. Doing so we get
\begin{equation}\label{two}
\frac{c_s \eta_{rec}}{\eta_0} \sim 4 \pi \epsilon
\end{equation}
where $\epsilon$ is the inflationary slow roll parameter. The very simple quantum argument underlying this guess is to identify $\Delta(E)$ formally introduced above with $\epsilon$. In average the former expression implies that the location of the peaks are determined by (\ref{peaks}) with 
\begin{equation}\label{basic}
\rho \sim 4 \pi \epsilon
\end{equation}

Note that this information argument is not determining the heights of the peaks\footnote{It is self-evident that one can construct inflationary models that violate~\ref{basic}. This in itself highlights the predictive power of our description of the early universe. }.

\begin{figure}
\vspace*{1cm}
\includegraphics[width=\columnwidth]{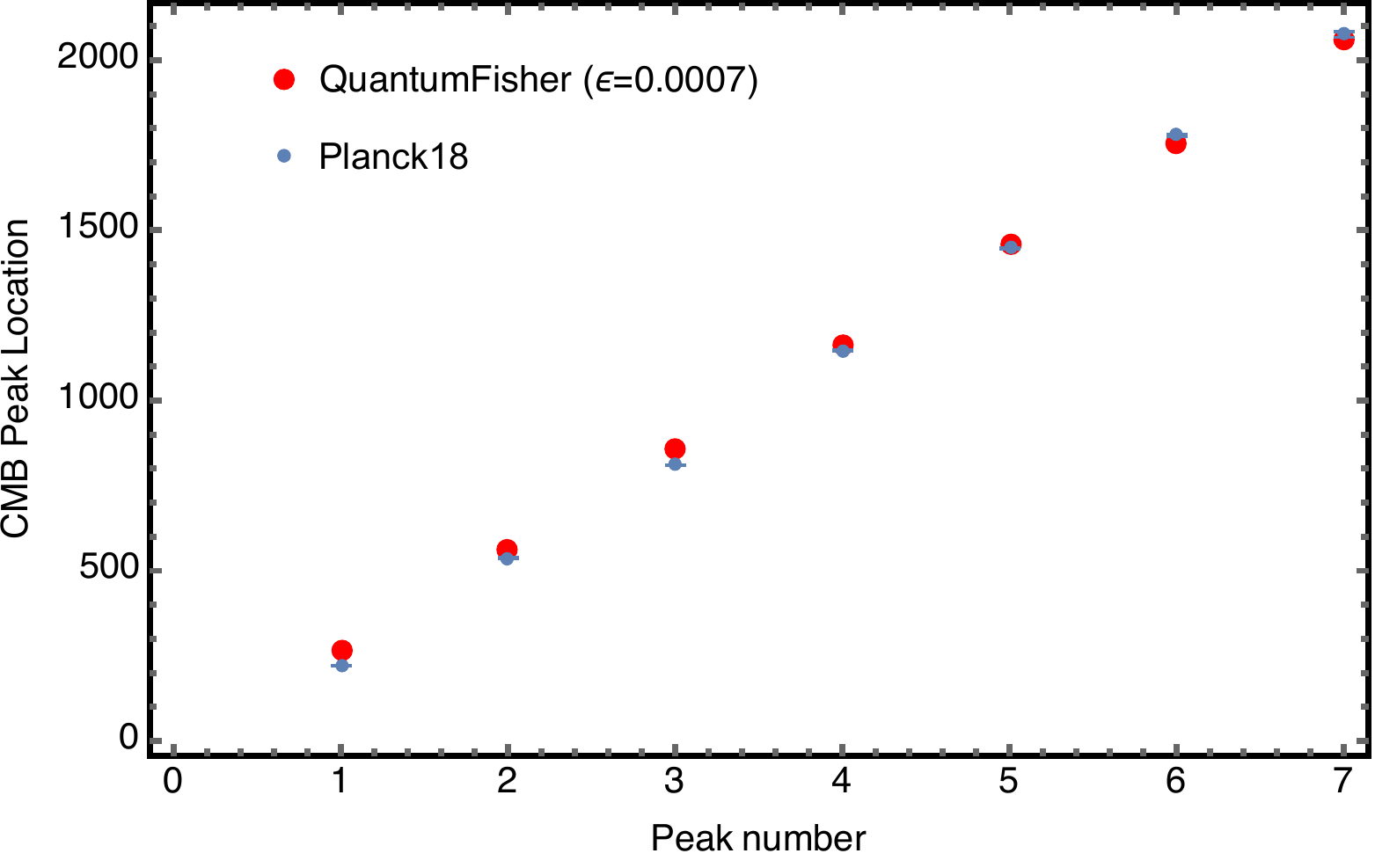}
\caption{Position of the first seven peaks in the CMB temperature power spectrum as measured by Planck18 (blue dots with corresponding error bars). We overplot the fitting model from quantum information theory (red dots). The allowed range for fitting the data in $\epsilon$ is 20\%. $\epsilon = 0.0007 \pm 0.00014$. This translates into $r = 0.01 \pm 0.002$.}
\label{fig:models}
\end{figure}

Unfortunately, in the region of multipoles corresponding to the acoustic peaks we don't have an analytical expression for $\rho$ so we need to use a fit \cite{cmbslow} that encodes the combined information about density of baryonic matter, dark matter as well as the factor of projection defined by the dark energy. Using this fit we get
\begin{equation}
\rho = 0.0101507
\end{equation}
Using~(\ref{basic}) we can estimate the allowed range of $\epsilon$ by compering it to the location of the temperature peaks as measured by Planck18~\cite{Planck18}. Fig.~1 shows the comparison; would agreement is obtained for a value of $\epsilon = 0.0007$. Note that variations of this value by more than 20\% give a significant deviation from the observed Planck18 data.

%corresponding to 
%\begin{equation}
%\epsilon \sim O(10^{-3})
%\end{equation}
%We find this reasonable estimate for $\epsilon$ already very encouraging.
%
%\RJ{The position of the temperature CMB peaks can be written as
%\begin{equation}
%l_n = \pi \rho^{-1} (n - \frac{1}{8},
%\end{equation} 
%which is eq.~(101) in Ref.~\cite{cmbslow}. This equation has corrections of $\sim 10\%$ depending of whether the peaks are odd or even, but this level of accuracy is not relevant to our argument.  Using~\ref{basic} we can estimate the allowed range of $\epsilon$ by comperating it to the location of the temperature peaks as measured by Planck18~\cite{Planck18}. Fig.~1 show the comparison; would agreement is obtained for a value of $\epsilon = 0.0007$. Note that variations of this value by more than 20\% give a significant deviation from the observed Planck18 data. 

The former constraint on $\epsilon$ coming from the CMB data allows us to make a prediction on the tensor to scalar ratio. If we assume the consistency relation $r = 16 \epsilon$~\cite{consistency}, we obtain that our predicted value is $r=0.01  \pm 0.002$ (the current experimental bound is $r < 0.07$~\cite{Planck18}). Thus if our interpretation of the early universe is correct, primordial gravitational waves should be measured at that value in the near future.

Finally let us point out that for a fixed value of $c_s$ increase of the density of cold dark matter leads to lower the multipole position of the peaks. From the former expression this suggest that increasing the primordial value of $\epsilon$ implies an increase of the density of cold dark matter and reciprocally that a potential lower bound on the value of $\epsilon$ implies non vanishing cold dark matter density \footnote{This opens the interesting possibility of a deep connection between the existence of dark matter and the absence of eternal inflation.}.

In summary the basic equation (\ref{basic}), if we take it seriously, is telling us how the primordial quantum Fisher information is already imposing a constraint on the time of recombination and potentially on the budget distribution of the energy we observe. In more concrete terms the CMB acoustic peaks are simply reflecting the quantum distance between inflation and recombination \footnote{In our model we are taking as the path in Hilbert space a geodesic with constant value of the primordial Fisher function. In complexity language we could say that the Universe uses the less complex quantum path to achieve recombination.}. Quantum mechanically this distance is bounded by the primordial Fisher information and therefore is determined by the early story of the Universe.

\acknowledgments
We thank Slava Mukhanov for advise and critical comments, and Licia Verde and the anonymous referee for many useful comments. The work of CG was supported by grants SEV-2016-0597, FPA2015-65480-P and PGC2018-095976-B-C21. The work of RJ is supported by grant PGC2018-098866-B-I00.

\end{document}